\documentclass[twocolumn,prl,showpacs,superscriptaddress]{revtex4}
\usepackage{graphicx} 
\usepackage{amsmath} 
\usepackage{amssymb}
\usepackage{amsbsy}
\usepackage[ansinew]{inputenc} 

\newcommand{\vect}[1]{\boldsymbol{#1}}
\newcommand{\Tonset}{T_{\mbox{\scriptsize onset}}}

\begin{document}

\title{Dynamical theory of superfluidity in one dimension}
\author{Thomas Eggel}
\affiliation{Institute for Solid State Physics, University of Tokyo,
Kashiwa 277-8581, Japan}
\author{Miguel A. Cazalilla}
\affiliation{Centro de Fisica de Materiales CSIC-UPV/EHU. Paseo Manuel
de Lardizabal 5,  E-20018 San Sebastian, Spain}
\affiliation{Donostia International Physics Center (DIPC), 
 Manuel de Lardizabal 4, E-20018 San
Sebastian, Spain} 
\author{Masaki Oshikawa}
\affiliation{Institute for Solid State Physics, University of Tokyo,
Kashiwa 277-8581, Japan}
\begin{abstract}
A theory accounting for the dynamical aspects of the superfluid response of
one dimensional (1D) quantum fluids is reported.  In long 1D systems 
the onset of superfluidity is related to the dynamical suppression of quantum phase slips at
low temperatures. The effect of this suppression as a function of frequency 
and  temperature  is discussed within the framework of the relevant correlation function
that is accessible experimentally, namely the momentum response function. 
Application of these results  to the understanding of 
the superfluid properties  of helium confined in nanometer-size pores, 
edge dislocations in solid $^4$He,  and ultra-cold atomic gases is also briefly discussed.
\end{abstract}
\date{\today} \maketitle

Superfluidity and superconductivity are often associated with the existence 
of long range phase coherence in quantum fluids.
Nevertheless, long range phase coherence is not a necessary condition for 
superflow, as the observation of superfluid response in two
dimensions (2D)~\cite{Reppy} demonstrates.  In 2D $^4$He films,
torsional oscillator (TO) experiments have established~\cite{Reppy}
existence of superfluidity (observed as a change in the resonance
frequency of the oscillator), despite lack of the long-range order.
The phase transition to superfluid phase is
related to the binding of (free ranging) vortices and anti-vortices into
pairs, as described by Berezinskii, Kosterlitz, and Thouless~\cite{BKT}. 
The 2D superfluidity without long-range off-diagonal order
can still be understood in terms of the helicity
modulus~\cite{Helicitymod},
which is a
thermodynamic (\emph{i.e.} static) property.
However, superfluidity manifests itself  
experimentally as a {\em dynamical} property and
in 2D, dynamical corrections~\cite{AHNS} 
to the helicity   modulus are important in understanding 
experimental observations.
In one dimension (1D), the dynamical aspect is even more important,
since the helicity modulus vanishes altogether in the thermodynamic
limit~\footnote{see supplementary material available at}.
That is, dynamical effects are not just corrections
to the static picture, but are key to the understanding of superfluidity
in 1D.

Recent TO experiments have detected superfluidity in long ($0.2-0.5\, \mu\mathrm{m}$)
nanometer-sized pores filled with liquid $^4$He~\cite{Toda07,Taniguchi10}, where a suppression of the 
superfluid onset temperature by pressurization and reduction of the pore diameter was observed. 
In optical lattices, it was found that a Bose-Einstein condensate of 
ultracold $^{87}$Rb atoms exhibits coherent current oscillations~\cite{Cataliotti01,Fertig05}. However,
when confined to 1D, the motion of the same ultracold degenerate gas becomes strongly
damped even in the presence of a relatively weak periodic potential~\cite{Fertig05}. 
In the case of supersolid $^4$He~\cite{Supersolid} it has been suggested theoretically~\cite{Boninsegni07,Shevchenko09} 
and experimentally~\cite{BalibarNature} that one likely explanation for the observations is
related to the superfluid properties of  edge dislocations in solid helium, which, as shown by Quantum 
Monte Carlo simulation~\cite{Boninsegni07}, behave as 1D quantum fluids~\cite{Haldane81,Cazalilla,TGbook}. 
These observations calls for a careful analysis of the notion of
superfluidity in 1D, despite the absence of helicity modulus in the
thermodynamic limit.  
The superfluidity in 1D is an essentially dynamical phenomenon, which
also reflects peculiarities of dynamics in 1D.
Indeed, compared to 2D and 3D,
dynamics in 1D tends to be much more constrained by the existence of   
conserved quantities.
Recently, this has been shown  to 
prevent complete thermalization~\cite{quench} or the total decay of a
current~\cite{Zotos} in 1D integrable systems.

In higher dimensions,
decay of superflow is caused by the motion of quantized vortices 
perpendicular to the direction of the flow.  In 1D, such a phenomenon 
corresponds to the creation of a
topological excitation, namely a phase slip (PS), which `unwinds'
the phase difference imposed upon the system and whose importance 
for the dynamical aspects of superfluidity in 1D 
has been pointed out by several authors. In the framework of Landau-Ginzburg
theory, a first calculation of the thermal production rate of PS  was given in Ref.~\cite{LangerAmbegaokar}.
Later, these calculations have been extended to the  quantum regime~\cite{Klebnikov, Shevchenko09}.
In homogenous systems,  the PS production rate is exponentially small at
low temperatures, implying that the lifetime of the superflow in 1D can be \emph{astronomically} long.
However, understanding of the suppression of superfluidity in the experiments mentioned
above~\cite{Taniguchi10,Fertig05} would require a finite PS production rate
even at low temperatures. Moreover, the connection of the PS production rate
to the experimental signatures of superfluidity, such as the response of a 
TO, remains obscure.
 
\begin{figure}[h]
 \centering
\includegraphics[width=0.4\textwidth]{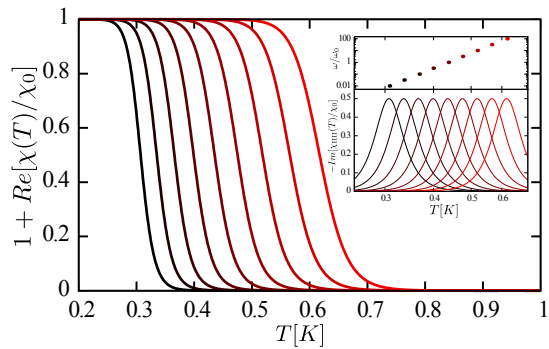}
\caption{(color online) 
Superfluid response as the frequency ($\omega_0= 2$kHz~\cite{Taniguchi10}). We used 
two terms for $H_{QPS}$ (cf. Eq.~\eqref{ham:qps}) setting $\Delta k_{10} = 0.007 a^{-1}_{0}$ and $\Delta k_{11}/\Delta k_{01} = 0.7$. 
$\chi_0 = M^2 v K /\pi \hbar$. The inset  shows the imaginary parts of $\chi(\omega)$ (lower part) and  the $\omega$-dependence of the dissipation peak temperature  on a log-log scale. For these parameters, the dependence is (roughly) a power-law of $\omega$ 
following from the power-law prefactor of the memory matrix (see supplementary material). 
Linear density and cut-off were chosen so as to conform to the experimental situation in \cite{Taniguchi10}, and we have taken the velocity of sound $v=200$m/s  and the Luttinger parameter $K=$8.1, resulting in onsets of superfluidity that agree well with experimental observations.}
\label{pic:omega_dep}
\end{figure}

\begin{figure}[h]
   \centering
\includegraphics[width=0.44 \textwidth]{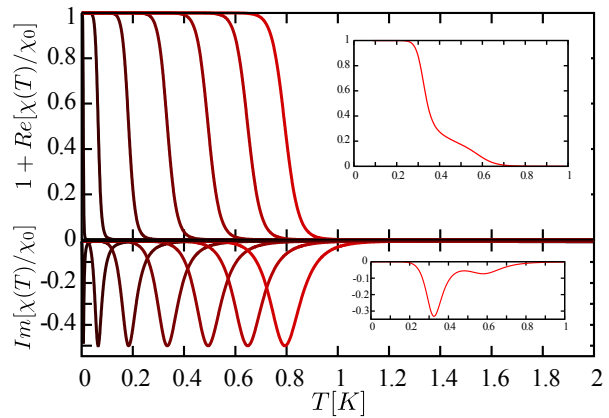}
\caption{(color online)  
Superfluidity response as a function of the the Luttinger parameter $K$, which is determines the compressibility of the 1D fluid.
 $K$ ranges from 3.2 (dark) to 10.2(bright)  ($\Delta k_{10} = 0.001 a^{-1}_0$, and $\Delta k_{11}/\Delta k_{10}  = 0.7$).
The insets ($K=6.2$) show the effect of a larger value of the PS momenta ($\Delta k_{10} = 0.5a^{-1}_0$ and $\Delta k_{11}/\Delta k_{10}  = 0.7$),  resulting in the appearance  of two dissipation peaks. } 
\label{pic:klutt_dep}
\end{figure}

In this Letter, we develop a  theory  of superfluidity in 1D which 
emphasizes the  dynamical aspects, as 
experimentally superfluid properties are  probed at finite frequencies and
are related to the dynamical momentum response function.
To compute the momentum response we used the memory matrix formalism, which
allows for a perturbative treatment in the operators describing the quantum PS. 
By analyzing the
momentum response assuming a periodic potential (which is a relevant model
for the $^4$He systems of Refs.~\cite{Taniguchi10,Boninsegni07} and the 1D ultracold atomic gases in optical 
lattice~\cite{Fertig05,Tokuno10}), we show that the superfluid onset temperature decreases with 
decreasing  the probe frequency (cf. Fig.~\ref{pic:omega_dep})  or decreasing 
 the compressibility of the fluid (cf. Fig.~\ref{pic:klutt_dep}). The latter 
can provide an explanation for the pressure-dependent suppression  
of superfluidity observed in the nanopore experiments of Ref.~\cite{Taniguchi10}.
 
Let us recall the description of superfluidity in terms of the 
momentum response function~\cite{Baym}, which 
is the response of the system to the motion of 
the container walls. In higher dimensions, the fluid-wall interaction 
affects only the atoms in the neighborhood of the container, 
and  it can be replaced by an appropriate boundary condition on the wall velocity field~\cite{Baym}.
The fraction of the fluid that is dragged along by a slowly moving container is
given by the transverse part of the static  momentum response function~\cite{Baym} which is defined as follows:  
Let $\vect{\Pi}(\vect{r})=\frac{\hbar}{2i} \left[\Psi^{\dag}(\vect{r})\vect{\nabla} 
\Psi(\vect{r}) -  \vect{\nabla} \Psi^{\dag}(\vect{r}) \Psi(\vect{r}) \right]$ be the momentum current operator and 
$\chi_{\mu\nu}(\vect{r},t)=  -i\hbar^{-1} \vartheta(t) \langle \left[\Pi_{\mu}(\vect{r},t), \Pi_{\nu}(\vect{0},0) \right]\rangle$ ($\mu,\nu=x,y,z$)
its response function, whose Fourier transform is a rank-2 tensor and for an isotropic fluid 
$\chi_{\mu\nu}(\vect{q},\omega) = (\delta_{\mu\nu} -
\frac{q_\mu q_{\nu}}{q^2}) \chi_T(q,\omega) + \frac{q_\mu q_{\nu}}{q^2}
\chi_L(q,\omega)$, where $\chi_{T(L)}(q,\omega)$ is the transverse
(longitudinal) momentum current response. The normal
component density is $\rho_n = -\frac{1}{M}\lim_{q\to0} \lim_{\omega\to 0}
\chi_T(q,\omega)$~\cite{Baym,convention} ($M$ is the particle mass).
In contrast, in 1D, the container wall
affects the entire fluid.
Therefore, its effect cannot be replaced by a boundary condition and
has to be explicitly accounted for 
in the calculation of the momentum response.
In fact, in 1D, $\chi(q,\omega)$
is a scalar which prevents the separation in a transverse and a longitudinal part.

With an eye on the experiments~\cite{Taniguchi10,Fertig05,Boninsegni07,BalibarNature},
we shall take a periodic potential to represent the container wall.
In the experiments of Ref.~\cite{Taniguchi10}, the walls of the pore are covered
by an inert layer of solid helium, which may be regarded as
periodic (allowing for 
a disorder potential is straightforward~\cite{TGbook} and will not alter our conclusions substantially). 
Furthermore, since we are interested in the low-temperature transport properties, we
shall rely upon the Tomonaga-Luttinger 
liquid (TLL) description of  1D fluids~\cite{Haldane81,Cazalilla,TGbook,Affleck,Affleck1} where
 the low temperature/frequency degrees of freedom of the
system are described by two collective (canonically conjugate) fields,
$\theta(x,t)$ and $\partial_x \phi(x,t)/\pi$, which account for phase
and density fluctuations, respectively. The effective Hamiltonian takes
the form $H = H_{0} + H_{\mathrm irr}$, where
\begin{equation}
H_0 = \frac{\hbar v}{2\pi} \int dx \left[ K (\partial_x \theta(x))^2 +
K^{-1} \left(\partial_x \phi(x) \right)^2 \right].
\label{eq:TLL}
\end{equation}
$H_0$ describes the properties of the system at $T = 0$, 
a 1D fluid of compressibility $K/(\hbar \pi v \rho^2_0)$ and sound velocity
$v$. Using the TLL Hamiltonian~\eqref{eq:TLL}, we find that
$\chi(\omega)= \lim_{q\to} \chi(q,\omega) = 0$, implying there is 
no normal component  and the system behaves as 
a perfect superfluid at any frequency and temperature.  Physically, this is because 
the Hamiltonian of Eq.~\eqref{eq:TLL} neglects the existence of quantum PS. 
This is, of course, a special property of ~\eqref{eq:TLL}, and the actual Hamiltonian of a 1D fluid 
involves  an infinite number of \emph{irrelevant} (in the renormalization group sense) operators 
$H_{\mathrm irr}$  and for the discussion that follows we
shall specialize to a particular subset of them, yielding the leading
corrections to $H_0$: 
\begin{equation}
H_{QPS} = \sum_{n>0,m}\frac{\hbar v g_{nm}}{\pi a^2_0}\int dx \cos
\left(2n\phi(x) + 2 \Delta k_{nm} x \right),\label{ham:qps}
\end{equation}
which describes the effect of quantum PS, 
responsible for the decay of the momentum current; $g_{mn}$ are 
dimensionless couplings related to the strength of the periodic potential and the interatomic 
interactions; $a_0\sim \rho^{-1}_0$ is a short-distance cut-off;  
$\hbar \Delta k_{mn} = (2 \hbar \pi \rho_0 - 2 m G) \hbar$ are the set of 
all  possible (lattice) momenta carried by the
PS ($\rho_0$ being the fluid's linear density). As mentioned above,
1D fluid is assumed to move in a  periodic background characterized by a minimum wave number $G$  
and the smallest $|\Delta k_{mn}|$ provides us with a measure of the incommensurability between  
the 1D fluid density and the wall potential~\cite{Haldane81,TGbook}.
For Galilean invariant systems,  $G = 0$ and $g_{n,m\neq0} =0$ and $v K 
= v_F= \hbar \pi \rho_0/M$. Irrelevant terms like  $a^{2(n+m)}_0 \int dx \: \left( \partial_x
\phi\right)^{2n} \left(\partial_x \theta \right)^{2m}$ ($m+n > 2$), etc., 
accounting for the curvature  of the phonon dispersion, do not, to leading
order, contribute to the decay of the momentum current.
However, the leading irrelevant correction, $H^{\prime}_{irr} = \frac{\hbar v K}{2\pi^2 \rho_0}
\int  dx \: \left( \partial_x \phi \right) (\partial_x\theta)^2$~\cite{Affleck1}, will be taken into
account below when obtaining the low-energy form of momentum operator, $\Pi$.

 In order to obtain the expression of the momentum operator at low temperatures/frequencies,  
we make a (time-dependent) unitary transformation to a frame where the walls are at rest~\cite{Tokuno10}.
This renders the calculation of the momentum response akin to the response of the system
to an external gauge field proportional to the velocity of the walls $v(t)$~\cite{Baym,Tokuno10}. 
Thus, $\Pi(x,t) = M j(x,t)$, where $j(x,t)$ is the particle current operator. 
From the continuity equation, $\partial_t \rho(x,t) + \partial_x j(x,t)$,
and $\rho(x) \simeq \rho_0 + \frac{1}{\pi} \partial_x\phi(x)$~\cite{Haldane81,TGbook,Cazalilla}, to leading
order, $j(x,t) = - \frac{1}{\pi}\partial_t \phi(x,t) = (i\pi \hbar)^{-1} \left[ H_0+ H_{irr}, \phi(x,t) \right]$~\cite{TGbook}.  
 Hence,  including the leading order
correction, $\Pi \simeq  - \frac{M}{\pi}  \int dx\, \partial_t \phi(x,t)  
 = \int dx \left[ \frac{M v K}{\pi}   + \frac{M v K}{\pi \rho_0} \partial_x \phi(x) 
\right] \partial_x\theta(x) =  J + \frac{v K}{v_F}P$, 
where $J = \frac{M v K}{\pi} \int dx\, \partial_x \theta(x)$ is the total particle (mass) current  
and $P = \frac{\hbar}{\pi} \int dx \, \partial_x\phi(x) \partial_y \theta(x)$ total energy current. 
Note that $\Pi$ is a linear combination of $J$ and $P$, and these two operators
are independently conserved by $H_0$ (\emph{i.e.} $[H_0,J] = [H_0,P]=0$ and $\langle J P \rangle_0 = 0$,
where $\langle\ldots\rangle_0$ is taken with respect to $H_0$).
However, since both $J$ and $P$ do not commute with $H_{QPS}$, in the presence of PS they become
dynamically coupled  and will acquire different decay rates.  
These effects can be taken into account  within the memory matrix formalism~\cite{Forster,Giamarchi91,RoschAndrei},
which has been successfully employed to compute the AC conductivity $\sigma(\omega)$ of charged 
1D systems~\cite{Giamarchi91,RoschAndrei}. 

In terms of the memory matrix $M(\omega;T)$ the momentum response can be written as
\begin{equation}
\chi(\omega;T) = \mathrm{Tr} \left\{ V \left[
 \omega \vect{1}  + i M(\omega;T)\right]^{-1} i
M(\omega;T) \boldsymbol{\chi}(T)\right\}
\label{eq:momresp}
\end{equation}
where $V_{ij} =   (v K/v_F)^{i+j-2}$ ($i,j=1,2$) and
$\boldsymbol{\chi}(T)$ is the matrix of 
static susceptibilities (see supplementary material for detailed definitions).
 In Fig.~\ref{pic:omega_dep} we have plotted the real and imaginary parts of the 
 momentum response $\chi(\omega)$ against the absolute temperature,
 for different values of the probe frequency. In the inset we show the  dissipation 
peak positions as a function of  the probe frequency. 
The parameters of the system (see caption for more details) 
 are chosen so as to reproduce   onset temperatures comparable to those experimentally observed in 
 liquid $^4$He filled nanopores of Ref.~\cite{Taniguchi10} when the probe 
 frequency equals $2$kHz~\cite{Taniguchi10}. As the probe frequency is decreased (corresponding to 
 darker colored curves), the onset temperature decreases. Indeed, this behavior can be 
 anticipated by taking the limit of $\omega\to 0^{+}$ in \eqref{eq:momresp}, which yields 
 $\chi(\omega\to 0, T) = \mathrm{Tr}\, \left[ V  \boldsymbol{\chi}(T)
 \right] = - \frac{ M^2 v K}{\hbar \pi} 
  - \left(\frac{vK}{v_F}\right)^2\frac{\pi(k_B T^2)}{6\hbar v^3}$, which is in 
stark contrast with the vanishing result obtained by neglecting the PS. 
The limiting behavior at $\omega \rightarrow 0$ is also consistent with
the vanishing helicity modulus, namely absence of superfluidity in
static sense. On the other hand, highly constrained dynamics in 1D leads to
superfluidity observable even at very low frequency
such as $2$kHz in Ref.~\cite{Taniguchi10}.
 
 In Fig.~\ref{pic:klutt_dep} we show real and  imaginary parts of the momentum response for several 
 values of the TLL parameter $K$, which determines the compressibility of the fluid. The onset 
 temperature is suppressed as the compressibility decreases (i.e.  as strength 
 of the  the atom-atom increases). This is in agreement with the expectation that strong 
 interactions tend to suppress the superfluid response.  In the experiment of Ref.~\cite{Taniguchi10}
 the value of $K$ is expected to decrease as pressure is applied to the system. Thus, the results displayed in  
 Fig.~\ref{pic:klutt_dep} are consistent with the experimental observation that the 
 onset temperature is suppressed by pressurizing the sample. Note also that since two separate currents are taken into account 
in \eqref{eq:momresp}, depending on the parameters used, there exists the possibility of two dissipation peaks with comparable weight, as displayed in  the insets of Fig.~\ref{pic:klutt_dep}.

In ultracold atomic systems, the momentum
response can be probed as described in Ref.~\cite{Tokuno10}.
Thus, here we restrict ourselves to discussing how the momentum response is probed by 
a  torsional oscillator (TO).
Our starting point is Newton's equation of motion 
for a TO cell filled with liquid $^4$He:
\begin{equation}
\frac{d}{dt} \left[ L_z(t) + \mathcal{L}_z(t)\right] = - \kappa
\varphi(t) - \eta \dot{\varphi}(t) + \tau_{\mathrm{ext}}(t),
\label{eq:newton}
\end{equation}
where $L_z(t) = I_0 \dot{\varphi}(t)$ is the angular momentum and $I_0$
the moment of inertia of the empy TO, $\varphi(t)$ is the rotation
angle, $\dot{\varphi}(t) = d\varphi(t)/dt$, $\kappa$ is the restoring
torque per unit angle, $\eta$ the friction coefficient, and
$\tau_{\mathrm{ext}}(t)$ the external torque driving the
TO. $\mathcal{L}_z(t)$ is the angular momentum of the \emph{normal}
component of the helium sample (which is dragged along with the TO). Quantum
mechanically, $\mathcal{L}_z(t) = \int d\mathbf{r} (\mathbf{\hat{z}}
\times \mathbf{r}) \cdot \langle \boldsymbol{\Pi}(\mathbf{r},t)\rangle$.  
For low rotation frequencies, this quantity can be computed within linear response
theory: $\langle \Pi_{\mu}(\mathbf{r},t)\rangle = - \sum_{\nu}\int dt d\mathbf{r}^{\prime}\, 
\chi_{\mu\nu}(\mathbf{r},\mathbf{r}^{\prime},t-t^\prime)\: ( \dot{\varphi}(t^\prime) \mathbf{\hat{z}}\times \mathbf{r}^{\prime})$
where    $\chi_{\mu \nu}(\mathbf{r},\mathbf{r}^\prime,t;T)$ is the  momentum response of the
liquid $^4$He in the TO cell  at an absolute temperature $T$~\cite{convention}. Hence,
$\mathcal{L}_z(t) = - \int dt \, \chi_{n}(t-t^\prime;T)\:
 \dot{\varphi}(t^\prime)$ where  $\chi_{n}(t;T) = \sum_{\mu,\nu}\int d\mathbf{r}
d\mathbf{r}^{\prime} \, g_{\mu\nu}(\mathbf{r}, \mathbf{r}^{\prime})
\chi_{\mu\nu}(\mathbf{r},\mathbf{r}^\prime,t;T)$ ($\mu,\nu = x,y,z$), and
$g_{\mu\nu}(\mathbf{r},\mathbf{r}^{\prime}) = \left(\mathbf{\hat{z}}
\times \mathbf{r}\right)_{\mu} \left(\mathbf{\hat{z}} \times
\mathbf{r}^{\prime}\right)_{\nu}$.The TO response is defined by the relationship $\varphi(t) = -\int dt \,
\chi_{TO}(t-t^\prime) \tau_{\mathrm{ext}}(t^\prime)$.  Hence,  from \eqref{eq:newton},
\begin{equation}
\chi^{-1}_{TO}(\omega) =  \omega^2 \left[ I_0 - 
\chi_{n}(\omega; T) \right]  + i \eta \omega  - \kappa.
\label{eq:TOresp}
\end{equation}
In Ref.~\cite{Nussinov07}, the TO response has recently been discussed
on phenomenological grounds. The expression reported there is  
identical to our Eq.\eqref{eq:TOresp} provided we identify  $g(\omega;T) = - \omega^2
\chi_{n}(\omega;T)$ where $g(\omega;T)$ is the back action function
introduced in Ref.~\cite{Nussinov07}. From \eqref{eq:TOresp} we see that
the moment of inertia of the empty TO, $I_0$ is corrected by $\delta I_n = -\mathrm{Re} \: \chi_n(\omega;T)$
and the friction coefficient, is corrected by $\delta \eta_n = -\omega \mathrm{Im}\, \: \chi_n(\omega;T)$.
To relate $\chi_n(\omega;T)$ to the momentum response of a 1D fluid
computed above,  we  imagine a straight 1D channel 
located at $\mathbf{r}$, filled with liquid $^4$He, and oriented along the
unit vector $\mathbf{\hat{d}}$. The momentum flow of the normal
component is  $\langle \boldsymbol{\Pi}(\mathbf{r},t) \rangle \propto \mathbf{\hat{d}}$ and
the velocity of walls of the channel is $\mathbf{\hat{d}}\cdot\left(\mathbf{z}
\times \mathbf{r}\right) \dot{\varphi}(t)$. For a typical sample size and
TO oscillator driving frequencies ( $\sim 10^3$ Hz~\cite{Taniguchi10}) this velocity
field varies very slowly on the scale of the length of the channel ($\sim 0.3 \, \mu m$~\cite{Taniguchi10}). 
Furthermore, we assume that finite-size 
effects can be neglected~\footnote{At the lowest temperatures accessible in the experiments
of Ref.~\cite{Taniguchi10}, $T \sim 0.1$ K the thermal length $L_T
\simeq \hbar v_s/T\sim 10 \, \mathrm{nm} \ll L \sim 0.3 \mu m$, where $v_s \sim
10^2$m/s, as obtained from specific heat measurements~\cite{Toda07})}. Therefore, 
the momentum response of the channel is given by the 
$q=\mathbf{q}\cdot\mathbf{\hat{d}} \to 0$  limit of $\chi(q,\omega;T)$ computed for an
infinite 1D system. However, besides the liquid $^4$He filling the 1D channel,  there may be
an additional contribution to $\chi_n(\omega;T)$ from other sources (in the experiment of 
Ref.~\cite{Taniguchi10} these would correspond to liquid helium filling the cavities between the nanoporous 
pellets). We shall assume that, around the onset temperature, these contributions 
only provide a weakly temperature and frequency dependent background signal. Thus,
$\chi_{n}(\omega; T) \simeq G_0 \chi(\omega;T) + \mathrm{const.}$ where  $G_0$ is a geometrical factor that
measures the relative weight of the 1D channel network to the total
response of the sample in the TO cell. Hence, the change in frequency of the TO
is $\delta \omega(T)= \omega(T) - \omega_{+\infty} \sim G_0 \left[ 1 +  \mathrm{Re}\: 
\chi(\omega_0;T)/\chi_0 \right]$, where $\omega_0  = \sqrt{\kappa/I_0}$ and is the empty TO frequency 
(we neglect the difference between the TO resonance frequency and $\omega_0$ 
and $\omega_{+\infty} =  \sqrt{\kappa/(I_0 + G_0 \chi_0)}$, where 
$\chi_0 = - \lim_{T\to0} \lim_{\omega\to 0} \chi(\omega;T) = M^2 v K/\pi\hbar$.  
The change in the quality factor is $\delta Q(T) \sim -G_0 \: \mathrm{Im} \chi(\omega;T)$.  
Note that $G_0$ accounts for the distribution of orientations of the channels within the TO cell.   
Indeed, within the TO cell some 1D channels will  be oriented perpendicular to flow of the walls (that is,
$\mathbf{\hat{d}}\cdot\left(\mathbf{z} \times \mathbf{r} \right) \simeq 0$)
and will not contribute to the change in the moment of inertia. 

 Finally, let us briefly discuss the relevance of our results for the
observed supersolid behavior in solid
$^4$He~\cite{BalibarNature,Supersolid}.  It has been suggested that
 in samples consisting of single crystals an explanation of 
of the observed supersolidity~\cite{Supersolid,BalibarNature} are edge dislocations.
Indeed, associated with the onset of superfluidity in
this system, there is a prominent dissipation peak and a stiffening
of the crystals. The latter is believed to be related of the pinning of
dislocations by $^3$He impurities~\cite{BalibarNature}. At sufficiently
low temperatures, the pinned dislocations become straight and behave as 1D
superfluid channels~\cite{Boninsegni07} that can be described as
Tomonaga-Luttinger liquids with $K \simeq 5$~\footnote{Note that, in our
convention $K$ is $1/K$ in the convention of
Ref.~\cite{Boninsegni07}.}. The results reported here
are fully applicable to such systems  and improve on earlier theoretical
treatments~\cite{Shevchenko09,Boninsegni07}. Experimentally, it may be
also interesting to further investigate the similarities in the TO response 
between the nanopore  systems~\cite{Toda07,Taniguchi10} and
the supersolid behavior of $^4$He single crystals.

 To summarize, we have reported a dynamical theory of superfluidity in 
 1D quantum fluids,  showing that in 1D superfluidity  is essentially a dynamical phenomenon 
 related to the suppression of quantum PS at low temperatures. 
 Our calculations go beyond previous theoretical treatments by computing the experimentally accessible 
 dynamical momentum response of the 1D fluid which has been obtained in a particular case  
 using the memory matrix formalism, allowing us to take into account the 
 dynamical coupling between the  particle and energy currents.  We have also
 demonstrated the explicit relation between this response function to the measurable parameters of 
 the torsional oscillator. 

We thank M. Suzuki and J. Taniguchi for enlightening discussions
on their TO experiments with liquid $^4$He in 1D.
MAC gratefully acknowledges the hospitality of
ISSP (University of Tokyo) and financial support from 
Spanish MEC grant FIS2010-19609-C02-02. MAC and TE thank 
D.~W. Wang for his hospitality at NCTS (Taiwan).  
TE acknowledges support by a MEXT scholarship (Japan).
The present work was partially carried out at the Supercomputer
Center, ISSP, University of Tokyo.

\clearpage

\appendix
\section{Supplementary Material}

\emph{The helicity modulus}

The  helicity modulus~\cite{Helicitymod} is defined
as $\Upsilon(T) = \lim_{L\to +\infty} \Upsilon(T,L)$ where  
$\Upsilon (T,L) = L  \left(\partial^2 F(\varphi,L,T)/\partial \varphi^2\right)\big|_{\varphi=0}$, where
$F(\varphi,T,L)$ is the free energy computed with twisted boundary
conditions on the field operator: $\Psi(x+L) = e^{i\varphi} \: \Psi(x)$. For a 1D system
described by the TLL Hamiltonian $H_0$ it holds that 

\begin{equation}
\Upsilon(T,L) = \Upsilon_0 \left(1+ \frac{\epsilon_0}{k_B T}\frac{\vartheta^{\prime\prime}_3(0,e^{-2 \epsilon_0/k_BT})}
{\vartheta_3(0,e^{-2 \epsilon_0/k_BT})}\right)
\end{equation}
 where  $\vartheta_3(z,q)=\sum_{n=-\infty}^{+\infty} q^{n^2} \: e^{2niz} $
is a Jacobi  theta function, and $\vartheta^{\prime\prime}_3(z,q) = \frac{d^2\theta_3(z,q)}{dz^2}$,
$\Upsilon_0 = \hbar vK/\pi$ and  $\epsilon_0 = \hbar\pi v K/L$, where $vK$ is the  phase stiffness at $T=0$. 
$\Upsilon(L,T)$ displays
an onset at $T \sim \epsilon_0\sim L^{-1}$. 
Hence  in the infinite system size limit 
\begin{equation}
\Upsilon(T) = \lim_{L\to+\infty} \Upsilon(L,T) =  0
\end{equation}
excluding the possibility of a non-vanishing static superfluid density. 
For a pore of $R = 2.8$ nm and lenght $L = 0.3\, \mu\mathrm{m}$~\cite{Taniguchi10} , we find that 
when we subtract the inert layer on the inside of the pores
that does not contribute to the superflow~\cite{Toda07}, at most $T_{\mathrm{onset}} \simeq 0.18$ K,
which seems too low when compared to the experimentally measured
$\Tonset \simeq 0.65$ K at a pressure of about 0.9MPa~\cite{Taniguchi10}.

\emph{The memory matrix}

As defined in the main text, the momentum response can be written in the 
memory matrix formalism~\cite{Forster,Giamarchi91,RoschAndrei} in the following way: 
\begin{equation}
\chi(\omega;T) = \mathrm{Tr} \left\{ V \left[
 \omega 1  + i M(\omega;T)\right]^{-1} i
 M(\omega;T) \boldsymbol{\chi}(T)\right\},
\end{equation}
where $\boldsymbol{\chi}(T) \simeq \mathrm{diag}\{
\chi_{JJ} , \chi_{PP}(T) \} = 
\mathrm{diag}\{ -\frac{M^2 v K}{\hbar \pi},   -\frac{\pi (k_BT)^2}{6\hbar v^3} \}
$ is the matrix of static susceptibilities and the 
memory matrix, $M(\omega;T)$ computed to lowest order
in $H_{irr}$ takes the form 

\begin{equation}
M(\omega;T) \simeq
\sum_{n> 0,m} \frac{\hbar^2 g^2_{nm}}{4\pi^2 a^4_0} D_{nm}(\omega;T) U_{mn} \boldsymbol{\chi}^{-1}(T)
\end{equation}
where 
\begin{equation}
U_{mn} = \left(\begin{array}{cc}  \left(\frac{2 n M v K}{\hbar} \right)^2 & -  \frac{2 n M v K}{\hbar}  \Delta k_{mn}\\
-  \frac{2 n M v K}{\hbar}  \Delta k_{mn}  & (\Delta k_{mn})^2 \end{array} \right)
\end{equation}
and  
\begin{equation}
D_{nm}(\omega;T) =
\left[F_{mn}(\omega;T)-F_{mn}(\omega=0;T)\right]/(i\omega)
\end{equation}
where  
\begin{align}
F_{mn}(\omega) &= -i\hbar^{-1}\int dx \int^{+\infty}_0 dt \, 
e^{i \omega t - i \Delta k_{mn}x} \: C_n(x,t) \ \notag\\
&= - \frac{2  a^2 _0\sin \left( \pi n^2 K\right)}{\hbar v}  \left(\frac{2 \pi k_B T a_0}{\hbar  v} \right)^{2n^2 K-2} \notag \\
&\quad B\left( \frac{n^2 K}{2} - i \frac{\hbar (\omega + v  \Delta k_{nm})}{4\pi k_B T} , 1- n^2 K\right) 
  \notag \\
&\quad \quad  \times  B \left( \frac{n^2 K}{2} -i \frac{\hbar  (\omega- v  \Delta k_{nm})}{4\pi k_B T}, 1- n^2 K \right)
\end{align}
being $O_{n}(x,t) = e^{-2in\phi(x,t)}$, and 
$C_n(x,t) = \langle [O^{\dag}_n(x,t), O_n(0,0)\rangle_0$, and 
$B(x,y)$ is the Euler Beta function. 

\begin{thebibliography}{30}
%
\bibitem{Reppy} D.~J. Bishop and J. Reppy, Phys. Rev. Lett. {\bf 40},
1727 (1979).
%
\bibitem{BKT} 
V.~L. Berezinskii Sov. Phys. JETP {\bf 34}, 610 (1972);
J.~ M. Kosterlitz and D.~J.  Thouless,  J. Phys. C {\bf 6}, 1181 (1973). 
%
\bibitem{Helicitymod}
M.~E. Fisher, M. Barber, and D. Jasnow, Phys. Rev. A {\bf 8}, 1111 (1973). 
%
\bibitem{AHNS} 
V. Ambegaokar, B.~I. Halperin, D.~R. Nelson, and
E. Siggia, Phys. Rev. Lett. {\bf 40}, 783 (1978).
%
\bibitem{Toda07} 
R. Toda \emph{et al.}, Phys. Rev. Lett. {\bf 99}, 255301 (2007).
%
\bibitem{Taniguchi10} 
J. Taniguchi, Y. Aoki, and M. Suzuki, Phys. Rev. B {\bf 82}, 104509 (2010).
%
\bibitem{Cataliotti01}
F.~S. Cataliotti \emph{et al.}, Science {\bf 293}, 5531 (2001).
%
\bibitem{Fertig05}
C.~D. Fertig \emph{et al.}, Phys. Rev. Lett. {\bf 94}, 120403 (2005). 
%
\bibitem{Supersolid} 
E. Kim and M. H. W. Chan, Nature {\bf 427}, 225
(2004); Science {\bf 305}, 1941 (2004); A.~S. Rittner and J. Reppy,
Phys. Rev. Lett. {\bf 97}, 165301 (2006).
%
\bibitem{Boninsegni07} 
M. Boninsegni \emph{et al.} Phys. Rev. Lett. {\bf 99}, 035301 (2007).
%
\bibitem{Shevchenko09} D.~V. Fil and S.~I. Sevchenko, Phys. Rev. B {\bf
80}, 100501(R) (2009).
%
\bibitem{BalibarNature} S. Balibar, Nature (London) {\bf 464}, 176
(2010).  X. Rojas \emph{et al.}, Phys. Rev. Lett. {\bf 105}, 145302
(2010).
%
\bibitem{Haldane81}
 F. D. M. Haldane, Phys. Rev. Lett. {\bf 47}, 1840 (1981)
 %
 \bibitem{Cazalilla}
 M. A. Cazalilla, J. Phys. B {\bf 37}, S1 (2004); 
 M. A. Cazalilla \emph{et al.}, arXiv:1101.5337 (2010).
 %
 \bibitem{TGbook}
 T. Giamarchi, \emph{Quantum Physics in One dimension} (Clarendon Press, Oxford, 2004). 
%
\bibitem{quench} 
T. Kinoshita, D. S. Weiss, Nature (London) {\bf 440},  900 (2006).
M. Rigol \emph{et al.}, Phys. Rev. Lett. {\bf 98}, 050405 (2007);
M. A. Cazalilla, Phys. Rev. Lett. {\bf 97}, 156403 (2006).
%
\bibitem{Zotos} 
H. Castella, X. Zotos, and P. Prelovsek, Phys. Rev. Lett. {\bf 74}, 972 (1995).
X. Zotos, \emph{ibid} {\bf 82}, 1764 (1999).
%
\bibitem{LangerAmbegaokar} 
V. Ambegaokar and J. Langer, Phys. Rev. {\bf 164},
498 (1967)
%
\bibitem{Klebnikov} 
S. Khlebnikov, Phys.~Rev.~Lett. {\bf 93}, 090403 (2004);
Phys. Rev. A {\bf  71}, 013602 (2005).
%
\bibitem{Tokuno10}
A. Tokuno and T. Giamarchi, report arxiv::1101.2469 (2010).
%
\bibitem{Baym} G. Baym, in \emph{ Mathematical Methods in Solid State
and Superfluid Theory}, edited by R. C. Clark and G. H. Derrick (Oliver
and Boyd, Edinburgh, 1969), p. 121.
%
\bibitem{convention}
The minus sign stems from the standard convention for retarded correlation functions 
(\emph{e.g.} A.~L. Fetter and J.~D. Walecka, \emph{Quantum Theory of Many-Particle Systems},
Dover Publications (New York, 2003). This convention differs by a minus sign from the one used 
in Ref.~\cite{Forster}.
%
\bibitem{Affleck1} 
A. del Maestro and I. Affleck, Physical Review B {\bf 82}, 060515(R) (2010).
%
\bibitem{Affleck}
A. del Maestro, M. Boninsegni, and I. Affleck, Phys.~Rev.~Lett. {\bf 106}, 105303 (2011).
11282 (2000).
%
\bibitem{Nussinov07} 
Z. Nussinov, A. V. Balatsky, M. J. Graf, and
S. A. Trugman, Phys. Rev. B {\bf 76}, 014530 (2007).
%
\bibitem{Forster} 
D. Forster,  \emph{Hydrodynamic fluctuations, broken symmetry, and correlation functions}, 
W. A. Bejamin (Reading, MA, 1975).
%
\bibitem{Giamarchi91}
T. Giamarchi, Phys. Rev. B 44, 2905 (1991).
%
\bibitem{RoschAndrei} A. Rosch and N. Andrei, Phys. Rev. Lett.{\bf 85},
1092 (2000)
%
\bibitem{unpub}
 T. Eggel, M. A. Cazalilla, and M. Oshikawa
(unpublished).
\bibitem{Yamashita}
K. Yamashita and D. S. Hirashima, Phys. Rev. B {\bf 79}, 014501 (2009).

\end{thebibliography}
\end{document}